\documentstyle[12pt]{article}
\begin{document}
\begin{titlepage}
\title{
{\bf Self-avoiding Gonihedric String \\
and \\
Spin  Systems}
}
{\bf
\author{
 G.K. Savvidy\\
 Physics Department, University of Crete \\
 71409 Iraklion, Crete, Greece\\
 and\\
 Institut f\"ur Theoretische Physik \\
 D-6000 Frankfurt am Main 11,Germany\vspace{1cm}\\
 K.G.Savvidy\\
 Moscow Institute of Physics and Technology\\
 141700 Moscow, Russia
}
}
\date{}
\maketitle
\begin{abstract}
\noindent

We classify different  theories of self-intersecting
random surfaces assigning special weights
to intersections. When self-intersection
coupling constant $\kappa$ tends to zero, then the surface
can freely inetrsect and it is completely self-avoiding
when $\kappa$ tends to infinity. Equivalent spin systems
for this general case were constructed. In two-dimension
the system with $\kappa = 0$ is in complete disorder
as it is in the case of 2D gauge Ising system.

\end{abstract}
\thispagestyle{empty}
\end{titlepage}
\pagestyle{empty}

{\Large{\bf 1.}}
\vspace{.5cm}

 In the recent article \cite{wegner1}
the authors formulated gonihedric string, which was introduced
in \cite{savvidy1,savvidy2,savvidy3},
on a euclidian lattice
and have found such spin systems with local interaction which
reproduce the same surface dynamics.

 In the case
of soft-self-avoiding surfaces they found, that the equivalent spin
system coinsides with usual or gauge Ising ferromagnet with
an additional diagonal antiferrormagnetic interaction. The coupling
constans of this spin systems are specially adjusted.
In the present article we shall
consider the more general case of self-avoiding and
non-self-avoiding surfaces.

 In a continuum euclidean space $R^{d}$
a random surface is associated with a connected polyhedral
surface $M$ with vertex coordinates
$X_{i}$, where $i = 1,..,\vert M \vert$ and $\vert M \vert$ is the
number
of the vertices. The energy functional for the polyhedral surface is
defined
as \cite{savvidy1}

$$A(M) = \sum_{<i,j>} \vert X_{i} - X_{j} \vert
\cdot \Theta(\alpha_{i,j}),\eqno(1)$$
where

$$\Theta(\pi) = 0 ,$$
$$ \Theta(2\pi -\alpha) = \Theta(\alpha),$$
$$\Theta(\alpha) \geq 0 ,\eqno(2)$$
the summation is over all edges $<i,j>$ of $M$ and $\alpha_{i,j}$
is the angle between two neighbor faces
of $M$ in $R^{d}$
having a common edge $<i,j>$.

 First we will classify different theories of
self-intersecting surfaces assigning special weights to
intersections \cite{savvidy3}. For non-self-avoiding surfaces it is
assumed that self-intersections do not
produce any additional energy and thus the surface can freely
intersect. In other cases self-intersections of a surface
produce an additional energy and one should define nontrivial
weights associated with these self-intersections. It was
suggested in \cite {savvidy3} to assign the following weights
to intersection

$$ \kappa  \sum_{common~edges} \vert X_{i} - X_{j} \vert
\cdot (\Theta(\alpha^{(1)}_{i,j}) +...
+\Theta(\alpha^{(r(2r-1))}_{i,j}) ),\eqno(3)$$
where $\kappa$ is self-intersection coupling constant,
$\alpha^{(1)}_{i,j},..,\alpha^{(r(2r-1))}_{i,j}$ are the
angles between the pairs of faces of $M$
at the common edge $<i,j>$ in $R^{d}$ in which $2r$ faces
intersect ~$r = 0,1,..,d-1$. It is natural to use $\kappa$
for the definition of the
different surface systems.

In euclidean space $R^{d}$ the coupling constant  $\kappa$
can be fixed from physical requirement. Indeed from continuity
principle \cite {savvidy3} it follows that
self-intersection coupling constant $\kappa$ should be

$$\kappa = \frac{1}{2} \eqno(4)$$
Physically this means that crumpled surface with folds can easily
tear into peaces in the places where the surface has acute angles.
Continuity principle guarantees that these transitions have non-zero
amplitude even at the $ \it classical~level$.

On the lattice in $Z^{d}$ the surface is
associated with the collection of
plaquettes, whose edgies are glued together pairwise and the
surface is closed if each edge of the lattice belongs to an
even number of plaquettes.
The angles between plaquettes are eighter $0, \pi /2~ or~\pi$.
There are no "continuous deformations" of the surface on the lattice
with respect to angles.
Therefore continuity principle does not tell us to much about the
way how to
choose $\kappa$ on the lattice and we
should keep $\kappa$ as a free parameter of the theory.

Thus one can distinguish different cases through the value of
$\kappa$.We will define three regions:

{}~~~~~~i)$\kappa =0$~- "bosonic" case

{}~~~~~ii) $0 < \kappa < \infty$~- "soft-fermionic" case

{}~~~~iii)$\kappa = \infty$~- "fermionic" case.

In the last case the surface in completely self-avoiding. In
\cite {wegner1} the authors considered the soft-fermionic
case $\kappa =1$.

Now we have a new
set of rules
(1) together with (3) for the energy functional
and therefore can define the surface
dynamics on the lattice with different self-intersection
coupling constant $\kappa$ .
The energy of the lattice surface can be
reexpressed through the plaquette variables $U_{P}$
($U_{P} = -1 $ if $P \in M$ and $U_{P} = +1$ if $P \not\in M$)
\cite{wegner1}. The energies of all possible eight configurations
at a given lattice edge in $Z^{3}$ can be expressed in the form

$$H^{\kappa}_{edge} = \frac{\kappa}{4}
(2-U_{1}-U_{-1})(2-U_{2}-U_{-2})
+ \frac{1-\kappa}{4} (2-U_{1}U_{-1}-U_{2}U_{-2})
 , \eqno(5)$$
where the given edge is in the 3rd direction and since only an
even number of plaquettes can meet at a given lattice edge,
the total product of corrresponding plaquette variables $U_{P}$
surrounding the edge is always equal to one
$U_{1}U_{-1}U_{2}U_{-2} = 1$. In (5) we have an extra term
which is zero in the case $\kappa = 1$ considered in
\cite{wegner1}.

In four-dimension, $Z^{4}$, again we have more complicated structure, than
in the case $\kappa = 1$. We find that

$$H^{\kappa}_{edge} =
\frac{\kappa}{4} (2-U_{1}-U_{-1})(2-U_{2}-U_{-2})$$
$$+\frac{\kappa}{4} (2-U_{2}-U_{-2})(2-U_{3}-U_{-3})$$
$$+\frac{\kappa}{4}(2-U_{3}-U_{-3})(2-U_{1}-U_{-1})  \eqno(6)$$
$$+ \frac{1-\kappa}{4}\left [ (U_{1} + U_{-1})(1-U_{2}U_{-2})
+ (U_{2} + U_{-2})(1-U_{3}U_{-3})
+ (U_{3} + U_{-3})(1-U_{1}U_{-1}) \right ]
. $$
The full energy associated with the lattice surface can always
be written as

$$H^{\kappa} = \sum_{over~all~adges}H^{\kappa}_{edge}\eqno(7)$$
Attaching spin variables to vertices of the dual lattice in
$Z^{3}$ and to links in $Z^{4}$ one can get the Hamiltonian in
terms of the independent spin variables $\sigma= \pm 1$ \cite
{wegner}. In $Z^{3}$ from (5) we get

$$H^{\kappa} = -2\kappa \sum_{\vec r,\vec \alpha}\sigma_{\vec r}
\sigma_{\vec r+\vec \alpha} +
\frac{1}{2}\kappa \sum_{\vec r,\vec \alpha,\vec \beta}
(\sigma_{\vec r}\sigma_{\vec r+\vec \alpha+\vec \beta} +
\sigma_{\vec r+\vec \alpha}\sigma_{\vec r+\vec
\beta})$$
$$-\frac{1-\kappa}{2} \sum_{\vec r,\vec \alpha,\vec \beta}
\sigma_{\vec r} \sigma_{\vec r+\vec \alpha}
\sigma_{\vec r+\vec \alpha+\vec \beta}
\sigma_{\vec r+\vec \beta}, \eqno(8)$$
where $\vec r$ is a three-dimensional vector whose components
are integer and $\vec \alpha$ ,$\vec \beta$
are unit vectors parallel to axis.
The Hamiltonian (8) is invariant under global $Z_{2}$
symmetry. But as it is easy to see from (8) the ground state  of
the system has larger degeneracy than
the ground state of the ferromagnet,
parallel layers of differently oriented spins have the same
energy. This has a strong influence on the
phase structure of the system.
High degeneracy of the ground state resembles gauge spin
systems with local $Z_{2}$ symmetry \cite{wegner}.

In $Z^{4}$ from (6) we get
$$H^{\kappa}= -\frac{5\kappa-1}{g^{2}} \sum_{plaquettes}
(\sigma\sigma\sigma\sigma) + \frac{\kappa}{4g^{2}} \cdot
\sum_{right~angle~plaquettes}
(\sigma\sigma\sigma\sigma_{\alpha})^{rt}
(\sigma_{\alpha}\sigma\sigma\sigma)$$
$$-\frac{1-\kappa}{4g^{2}} \cdot
\sum_{triples~of~right~angle~plaquettes}
(\sigma\sigma\sigma\sigma_{\alpha})^{rt}
(\sigma_{\alpha}\sigma\sigma\sigma_{\beta})^{rt}
(\sigma_{\beta}\sigma\sigma\sigma)  ,\eqno(9)$$
As it is seen from (8) and (9) the difference
with the case $\kappa=1$ \cite{wegner1}consists in the fact
that in (8) appears a new four-spin interaction
term and a three-plaquette interaction term in (9).
The partition function is defined as

$$Z(\beta) = \sum_{\{\sigma_{\vec r}\}} e^{-\beta H^{\kappa}}.
\eqno(10)$$
Expressions (5),(6) and (8),(9) together with (7) and (10)
completely define the desired system of random surfaces
on the lattice.
The advantage of the given above
formulation of the initial system of the
random surfaces as a spin system with local interaction
consists in the fact that now one can study more familiar
spin systems and their phase structure.
\vspace{1cm}

{\Large{\bf2.}}
\vspace{.5cm}

The systems of the same spirit can be formulated in any
dimension \cite{wegner1}. Below we will concentrate on the
two-dimensional
case where they present a model of random walks \cite {wegner1}.

The amplitudes of the elementary steps on the two-dimensional
square lattice can be represented in terms of the eight-vertex
model
\cite{baxter,fan,wu1,fanwu}
with the weights (see (5) with the correspondence:
$3D$~ edge~ $\longleftrightarrow$ ~ $2D$ ~ vertex, $3D$ ~ plaquettes
$\longleftrightarrow$ ~ $2D$ ~ bonds )

$$\omega_{1}=1,~~\omega_{2}=exp(-4\beta \kappa),~~
\omega_{3}=\omega_{4}=1,~~\omega_{5}=\omega_{6}=
\omega_{7}=\omega_{8}=exp(-\beta),\eqno(11)$$
where $\omega_{\xi}=exp(-\beta \epsilon_{\xi})$ and
$\epsilon_{\xi}$ is the energy assigned to the
$\xi$th type of vertex configuration
($\xi_{i}=1,..,8$)

$$\epsilon_{1}=0,~~ \epsilon_{2}=4\kappa,~~
\epsilon_{3}= \epsilon_{4}=0,~~ \epsilon_{5}=
\epsilon_{6}=\epsilon_{7}=\epsilon_{8}=1.\eqno(12)$$
The partition function (10) therefore becomes equal to
(see (5) and (7))

$$Z(\omega_{1},..,\omega_{8})=
\sum_{over~all~closed~paths}\prod^{N^{2}}_{i=1}~~
\omega_{\xi_{i}}\eqno(13)$$
where the summation is extended over all closed paths
consisting of vertices of even degrees (0,2 or 4) on
lattice with $N \times N$ sites. The evolution of the
partition function (13) is a
special case of more general problem involving the
counting of weighted graphs (paths) with 16
 kinds of vertex configurations
with arbitrary weights \cite {wu1}.
In our case (11) $\omega_{9}=...=\omega_{16}= 0.$

The partition function (13) posseses a number of evident
symmetries \cite {fan}

$$Z(\omega_{1},\omega_{2},
\omega_{3},\omega_{4};\omega_{5},\omega_{6},
\omega_{7},\omega_{8})=Z(\omega_{2},\omega_{1},
\omega_{4},\omega_{3};\omega_{6},\omega_{5},
\omega_{8},\omega_{7})$$
$$=Z(\omega_{4},\omega_{3},
\omega_{2},\omega_{1};\omega_{7},\omega_{8},
\omega_{5},\omega_{6})=Z(\omega_{3},\omega_{4},
\omega_{1},\omega_{2};\omega_{8},\omega_{7},
\omega_{6},\omega_{5})
.\eqno(14)$$
Less evident is the duality transformation of Nagle
\cite {nagle1,wu1,wegner2}. We shall write here
only the resulting linear transformation of the
eight weights (11) into 16 kinds of vertex
configurations appearing in the dual
representation \cite{wu1}

$$ \omega^{\star}_{\xi_{i}}=\sum^{8}_{\xi_{i}=1}~~
\omega_{\xi_{i}}~~\prod_{k}c_{i,k}(\xi_{i})
,\eqno(15)$$
where one should draw a bond between vertices $i$ and $k$
for each $c_{i,k}(\xi_{i})$

$$c_{i,k}(\xi_{i})=c_{k,i}(\xi_{i})=\left\{ \begin{array}{ll}
1  & \mbox{if $\xi_{i}$ has bond on the edge $<i,k>$} \\
-1 & \mbox{if $\xi_{i}$ does not have bond on the edge $<i,k>$}
\end{array}
\right.  \eqno (16)$$

The summation in (13) should be extended now
to all graphs which
include all 16 kinds of vertices of even
and odd degrees. It is easy to
compute these dual weights from (15) and (11)

$$a^{\star} = \omega^{\star}_{1}=\omega^{\star}_{2}=
4(1+\omega)-(1-\omega^{4k});~~~
b^{\star} = \omega^{\star}_{3}=\omega^{\star}_{4}=
4(1-\omega)-(1-\omega^{4k});$$
$$c^{\star} = \omega^{\star}_{5}=\omega^{\star}_{6}=
1-\omega^{4k};~d^{\star} = \omega^{\star}_{7}=
\omega^{\star}_{8}=1-\omega^{4k};~
\omega^{\star}_{9}=...=\omega^{\star}_{16}=1-\omega^{4k},
\eqno(17)$$
where $\omega = exp(-\beta)$. In the bosonic case i)$\kappa=0$
the dual system (17) reduces to the simple eight-vertex model

$$a^{\star}=1+\omega,~~ b^{\star}=1-\omega,~~c^{\star}=
d^{\star}=0
\eqno(18)$$
which can be easy solved \cite {fanwu}. The free energy
$f$ per vertex is

$$-\beta f = \ln a^{\star} =\ln \frac{a+b+c+d}{2}=
\ln (1+\omega),
\eqno(19)$$
where $a=\omega_{1}=\omega_{2}=1,~~ b=\omega_{3}=
\omega_{4}=1,~~c=\omega_{5}=\omega_{6}=\omega,~~
d=\omega_{7}=
\omega_{8}=\omega $. The free energy (19) is an analytic
function of $\beta$ and the system is completely desordered
(dual model (18) is in ferroellectric order
$a^{\star} > b^{\star}+c^{\star}+d^{\star}$). This is
a result of the high degeneracy of the ground state:
the vertices
$\omega_{1},~~\omega_{2},~~\omega_{3}$ and $\omega_{4}$ have the
same energy and the ground state is no longer unique and has
high degeneracy.
The transition temperature vanishes
as in the case of the gauge Ising model in
two-dimension \cite{wegner} or Ising model in one-dimension.

There are two large classes of the
general eight-verex model  which can be solved analitically:
$\alpha$)free-fermion model \cite{hurst,hurst1,gibberd,fan,fanwu}
,when

$$\omega_{1}\omega_{2}+\omega_{3}\omega_{4}=\omega_{5}
\omega_{6}+\omega_{7}\omega_{8} \eqno (20)$$
and $\beta$) zero-field  eihgt-vertex Baxter model
\cite{baxter}, when
$a=\omega_{1}=\omega_{2},~~ b=\omega_{3}=\omega_{4},~~
c=\omega_{5}=\omega_{6},~~ d=\omega_{7}=\omega_{8}$.
Nor of these conditions is fulfilled by the weights (11),
except of the case i)$ k=0$, which was already considered (18),(19).

To get information about the behavior of the general
system (11), when $ k\neq 0$, one can use fermionic representation
of the partition function (13).Following
\cite {hurst,hurst1,gibberd,fan,fanwu}
one can rewrite the partition function (13) as a product of
the fermion operators

$$Z(\omega_{1},..,\omega_{8})=
<0 \vert \prod^{N^{2}}_{i=1}
(\omega_{1} + \omega_{2}a^{+}_{i}b^{+}_{i}a_{i-N}b_{i-1}+
\omega_{3}a^{+}_{i}a_{i-N}+\omega_{4}b^{+}_{i}b_{i-1}+$$
$$+\omega_{5}a^{+}_{i}b_{i-1}+\omega_{6}b^{+}_{i}a_{i-N}
+\omega_{7}a^{+}_{i}b^{+}_{i}+\omega_{8}a_{i-N}b_{i-1}
\vert 0>
,\eqno(21)$$
where $a^{+}_{i},b^{+}_{i}$ and $a_{i},b_{i}$
are bond creation and annihilation
fermion operators at the i'th vertex,
$a$ refers to bonds in the vertical direction, $b$
to bonds in the horizontal direction and
$\vert 0 >$ is the vacuum
state. Using anticommutation relations
of the fermion operators one can get
\cite{hurst,hurst1,gibberd,fan,fanwu}

$$Z = \omega^{N^{2}}_{1}<0\vert~ T  exp \{ ~\sum^{N^{2}}_{i=1}
H(i)+g\cdot H_{int}(i) ~\} ~\vert 0> ,
\eqno(22)$$
where $H(i)$ is quadratic in the field operators

$$\omega_{1}H(i) = \omega_{3}a^{+}_{i}a_{i-N}+\omega_{4}b^{+}_{i}b_{i-1}+
\omega_{5}a^{+}_{i}b_{i-1}+\omega_{6}b^{+}_{i}a_{i-N}
+\omega_{7}a^{+}_{i}b^{+}_{i}+\omega_{8}a_{i-N}b_{i-1}
,\eqno(23a)$$
$H_{int}(i)$ is quartic

$$H_{int}(i) = a^{+}_{i}b^{+}_{i}a_{i-N}b_{i-1} \eqno(23b)$$
and

$$g=\frac{\omega_{1}\omega_{2}+\omega_{3}\omega_{4}-\omega_{5}
\omega_{6}-\omega_{7}\omega_{8}}{\omega^{2}_{1}}
\eqno (23c)$$
and $T$ orders the indices from $i=1$ to $i=N^{2}$.

For the system (11) quadratic part is equal to

$$ H(i) = a^{+}_{i}a_{i-N}+b^{+}_{i}b_{i-1}+
\omega (a^{+}_{i}b_{i-1}+b^{+}_{i}a_{i-N} +
a^{+}_{i}b^{+}_{i}+a_{i-N}b_{i-1})
\eqno(24a)$$
and the coupling constant (23c) is

$$g=1-2\omega^{2}+\omega^{4\kappa} \eqno (24c)$$
One can directly use field theoretical methods to evaluate
the expression (22). In high temperature regim when
$\beta \rightarrow 0$ the
coupling constant (24c) is small $g=4(1-\kappa)\beta +
4(2\kappa^{2}-1)\beta^{2}$, and one can use perturbation
expansion. In zero approximation we have

$$-\beta f_{0}= \frac{1}{N^{2}} \ln Z_{0}=
\frac{1}{N^{2}} \ln <0\vert ~Texp\{~\sum^{N^{2}}_{i=1}~H(i)~\}~ \vert 0>$$
$$= \frac{1}{8\pi^{2}} \int_{0}^{2\pi} \ln [(1-\omega^{2}+\omega^{4})
+(1-\omega^{2})(\cos\theta + \cos\theta \cos\phi +\cos \phi)]
d\theta d\phi \eqno(25)$$
The free egergy is analitic for all temperatures $\beta$
and system is always in disordered regime. The physical
reason of this type of phase structure of the system lies
in the fact,
that in (11) we don`t have dominant vertex, the
nessesary requirment for the order at low temperatures.
Of course this is a zero order result and as it is easy
to see from (24a) and (21) the expression (25) is reliable
only if~ $2\omega^{2}-1~ > ~0$~, that is for temperatures
$T$ greater than $\frac{1}{\ln \sqrt 2}$. It is also true
that at low tempetatures the coupling constant (24c) grows and
is of order one. Therefore more rigor analysis should be done
in the low temperature region to confirm zero order result
\cite{peierls,sinai}.

We expect more complicated phase structure of the system in
high dimensions where the energy functional has the meaning
of the size of the system and where it is not conformal invariant,
as it is in the two-dimension case (11).
Some exact results concerning three-dimensional edge models
have been obtained in \cite{nagle2,villain,orland,papanicolaou}

In conclusion we would like to stress that (9) can be
considered as an extention of the Wegner-Wilson action if one
exchange spin variables $\sigma$ by the elements
$U = exp(ig A)$ of $SU(N)$. The new action will contain
high derivative terms with specialy adjusted
coupling constants. This type of extension has been considered
before in \cite{weisz} with the aim to improve continuum limit
of the lattice gauge theory.

\vspace{1cm}

{\Large{\bf Acknowledgements}}
\vspace{.5cm}

One of the authors
(G.K.S.) is grateful to Alexander Belavin for the
discussion of the Wegner spin systems which took place in
Yerevan in 1978 and to Franz Wegner for numerous
stimulating conversations
and kind hospitality in Heidelberg.
We thank G.Athanasiu, E. Floratos, R. Flume,
N. Papanicolaou,
E. Paschos for interesting discussions and
W. Greiner for useful comments and hospitality
in Frankfurt University. One of the authors
(K.G.S.) is thankful to S.Pnevmatikos for kind
invitaton into Summer School of Heraklion.

This work was supported in part by the Alexander von
Humboldt Foundation.

\vfill
\end{document}